# Probing blue long persistent $Eu^{2+}/Nd^{3+}$ activated $(Ca_{0.9}Si_{0.1})Al_2O_6$ emerging phosphor based novel invisible ink for knocking out fake security papers and combat counterfeiting

Vishnu V. Jaiswal and D. Haranath[*]

*Luminescent Materials and Devices Group, Department of Physics, National Institute of Technology, Warangal – 506 004, Telangana, India*

**ABSTRACT**

In the current work, a novel class of $(Ca_{0.9}Si_{0.1})Al_2O_6:Eu^{2+},Nd^{3+}$ long persistent phosphor was synthesized using high-temperature (1000-1500°C) solid-state reaction method. Further, the phosphor material was formulated into a phosphorescent ink using epoxy based colloidal suspension. Interestingly the phosphorescent ink absorbs ambient energies such as white light from D65 and tungsten lamps, sunlight, artificial lights etc., stores up the energy and emit as efficient blue colored photoluminescence (PL) apparently visible for many hours for dark adapted human eye. However, the optimum PL emission centred at ~450 nm has been registered at ~365 nm excitation that is attributed to transitions from $4f^65d^1$ to $4f^7$ energy levels of $Eu^{2+}$ ions. The longer persistence of PL aroused due to trapping and de-trapping of holes at $Nd^{3+}$ sites situated proximate to valance band. In the optimised stoichiometric composition of the phosphor, the persistence lasted for >4 hours in extremely dark conditions. Furthermore, the phenomenon of long persistence enables to create/design unique hidden markings on goods and certificates using quick response (QR) code patterns. Systematic studies have been performed on various templates kept under UV (~365 nm) excitation to identify fake currencies, barcodes and combat counterfeiting effectively.

[*]Corresponding author. Tel.: +91 995 810 1115; Fax: +91 870 245 9547
*E-mail address*: haranath@nitw.ac.in (D HARANATH)

## 1. Introduction

Phosphor material having a special optical property of absorbing ambient energies such as roomlights, artificial lights, sunlight and/or ultraviolet (UV) and emit in another frequency of visible light for extended durations in dark after the source of excitation is turned-off is known as long-persistent (LP) phosphor. The persistent light emission is a result of trapping and de-trapping of either electrons or holes or both by metastable states or defect centers in the host lattice. As a result of which the recombination process is significantly delayed at room temperature (~20°C) (Blasse, 1968; Dorenbos, 2005; Lin *et al.*, 2001). The LP phosphor materials find numerous applications in the fields of dark vision displays, security codes and forensic exhibit detections, structural crack identifications, route markings, signages and in many more sectors (Kinoshita et al., 1999; Lisensky et al., 1996; Wu et al., 2017). The photoluminescence (PL) characteristics of these fascinating materials have been comprehensively studied in the above fields due to their exceptional quantum yields, dark persistence times and remarkable thermal stabilities (Bierwagen *et al.*, 2016; Xu *et al.*, 2014). In recent times, $Eu^{2+}/Nd^{3+}$ activated inorganic aluminate host lattices comprising group-IIA elements have gained enormous attention as an interesting class of efficient LP phosphors (Bonturim *et al.*, 2018; Haranath *et al.*, 2007; Jansen *et al.*, 1999; Wang and Wang, 2007). Further, these phosphors are extensively investigated as emergency signage indicators and dark vision display applications (Khattab *et al.*, 2019; Van den Eeckhout *et al.*, 2010). Conversely, another innovative class of phosphors based on silicoaluminate is less explored may be due to large variation of melting points of precursor chemicals. Remarkably silicoaluminate exhibit excellent chemical properties such as, efficient PL intensities, premier quantum yields, prolonged dark persistence times, substantial physico-chemical and aqueous stabilities (Matsuzawa, 2006; Zhang and Deng, 2017). The divalent Eu ion activated host lattices offer emissions covering entire visible (400-700 nm) region of the electromagnetic spectrum due to varied crystal-field effects offered by the host lattices (Poort *et al.*, 1995). Apart from that, the radiative transition occurs from excited state ($4f^65d^1$) to the ground state ($4f^75d^0$) resulting allowed dipole interaction (Swati *et al.*, 2015). The luminescent properties of LP phosphor could possibly be enhanced significantly by incorporating another rare-earth ion, $Nd^{3+}$ as co-activator. The effective combination of $Eu^{2+}/Nd^{3+}$ activated silicoaluminate phosphors unveil optimal PL due to $Eu^{2+}$ ions and longer persistence times associated with $Nd^{3+}$ ions in the host lattice (Aitasalo *et al.*, 2003; Cao *et al.*, 2019). Thus, these dominant features make $Eu^{2+}/Nd^{3+}$ activated calcium silicoaluminate LP phosphor as an emerging potential candidate in the field of security and currency counterfeit applications.

Further, to the best of our knowledge, there is no such report existing on the application of blue PL from $Eu^{2+}/Nd^{3+}$ activated calcium silicoaluminate (($Ca_{0.9-x}Si_{0.1}$)$Al_{2-y}O_6$:0.01$Eu^{2+}$,0.02$Nd^{3+}$ henceforth denoted as LDB4-LP) phosphor in the field of security code detection till now. The mere substitution of Si ion in the current host lattice not only endorsed decrement in energy band gap but also helped in enhancing absorption capabilities of the phosphor in the visible region. The LP phosphor based phosphorescent ink could be used for the security applications involving hidden confidential marking identification on quick response (QR) codes, bar codes and secure the anti-counterfeiting (Kanika *et al*., 2017; Krombholz *et al*., 2014; Meruga *et al*., 2015, 2012). For the design of various other LP phosphor materials, a high temperature solid-state reaction method employing inexpensive inorganic host lattices and suitable activator ions are the need of the hour.

In the current work, we have developed a novel compound of LP phosphor using a modest and cost-effective synthesis route executed under $H_2$ gas-free mild reducing atmosphere, which is merely reported in the literature. Additionally, the adapted synthesis route produces phosphor with high degree of uniformity, purity, yield, low-cost production with reliability for security applications. Suitable rare-earth doping and co-dopings offer a bright PL emission efficiency in the blue region and substantial LP effect. Thus, these results have established that LDB4-LP phosphor ink with prolonged persistence time represents a novel and beneficial class of LP ink that finds strategic applications in various sectors (Bite *et al*., 2018). Additionally, we have demonstrated promising application of LDB4-LP phosphor based transparent LP ink made from epoxy assisted polymeric solution for the visualization of latent markings on QR codes, product bar codes, and company identity cards to combat counterfeiting.

## 2. Experimental details

### 2.1. Synthesis of $(Ca_{0.9-x}Si_{0.1})Al_{2-y}O_6$:$xEu^{2+}$,$yNd^{3+}$ *(where x = 0 to 0.05 mol and y = 0 to 0.1 mol)*

The LP phosphor samples with stoichiometric composition $(Ca_{0.9-x}Si_{0.1})Al_{2-y}O_6:xEu^{2+},yNd^{3+}$ (where, x = 0 to 0.05 mol and y = 0 to 0.1 mol), were synthesized using high temperature solid-state reaction method in presence of $H_2$ gas-free mild reducing environment. Since x has been varied from 0 to 0.05 mol, the samples were numbered as LDB1, LDB2, LDB3, LDB4, LDB5, LDB6, LDB7 and LDB8 as listed in Table 1. The starting chemicals were used as calcium carbonate ($CaCO_3$; SRL, 99.9%), silicon dioxide ($SiO_2$; SRL, 99.9%) aluminum oxide ($Al_2O_3$; CDH, 99%), europium oxide ($Eu_2O_3$; Sigma-Aldrich, 99.9%), and neodymium oxide ($Nd_2O_3$; Sigma-Aldrich, 99.9%). The high temperature tubular furnace fitted with precision temperature regulator has been used for the synthesis of $(Ca_{0.9-x}Si_{0.1})Al_{2-y}O_6:xEu^{2+},yNd^{3+}$ (where, x = 0 to 0.05 mol and y = 0 to 0.1 mol) LP phosphors. The stoichiometric molar ratios (0.89:0.1:1.98:0.01:0.02 for Ca:Si:Al:Eu:Nd) of precursor chemicals were mixed homogeneously using mortar-pestle. The grounded powder was packed in re-crystallized alumina container and annealed under mild reducing $H_2$-free gas environment in the temperature range ~1000 to 1500°C for 1-5 hours in a tubular furnace. A steady ramp rate of 8±0.1°C/min was maintained for the furnace till it reached the set point. The synthesized sample has uniform white body color, which was grounded to powder form. The yield of the synthesized LP phosphor was more than 80% and maintained good homogeneity. The activator concentrations (x and y for $Eu^{2+}$ and $Nd^{3+}$) have been varied from 0 to 5 and 0 to 10 mol%, respectively in the stoichiometric composition of $(Ca_{0.9-x}Si_{0.1})Al_{2-2y}O_6:xEu^{2+},2yNd^{3+}$ host lattice. An optimal chemical composition of synthesized product ensures an excellent quality of initial intensity and decay times. The substitution of $Eu^{2+}$ and $Nd^{3+}$ take place by occupying $Ca^{2+}$ and $Al^{3+}$ sites, respectively. The optimized concentrations (x and y) were found to be at 1 and 2 mol%, respectively and labeled as sample number LDB4 in the listed Table 1.

**Table 1:** LP phosphor powder samples with nominal composition of $(Ca_{0.9-x}Si_{0.1})Al_{2-2y}O_6:xEu^{2+},2yNd^{3+}$.

| Sr. No. | $(Ca_{0.9-x}Si_{0.1})Al_{2-2y}O_6:xEu^{2+},2yNd^{3+}$ (x = 0 to 5 mol% and y = 0 to 10 mol%) | Sample Code |
|---|---|---|
| 1. | $(Ca_{0.90},Si_{0.1})Al_2O_6$ | LDB 1 |
| 2. | $(Ca_{0.8975}Si_{0.1})Al_{1.995}O_6:0.0025Eu^{2+},0.005Nd^{3+}$ | LDB 2 |
| 3. | $(Ca_{0.895}Si_{0.1})Al_{1.99}O_6:0.005Eu^{2+},0.01Nd^{3+}$ | LDB 3 |
| 4. | $(Ca_{0.89}Si_{0.1})Al_{1.98}O_6:0.01Eu^{2+},0.02Nd^{3+}$ | LDB 4 |
| 5. | $(Ca_{0.88}Si_{0.1})Al_{1.96}O_6:0.02Eu^{2+},0.04Nd^{3+}$ | LDB 5 |
| 6. | $(Ca_{0.87}Si_{0.1})Al_{1.94}O_4:0.03Eu^{2+},0.06Nd^{3+}$ | LDB 6 |
| 7. | $(Ca_{0.86}Si_{0.1})Al_{1.92}O_6:0.04Eu^{2+},0.08Nd^{3+}$ | LDB 7 |
| 8. | $(Ca_{0.85}Si_{0.1})Al_{1.90}O_6:0.05Eu^{2+},0.1Nd^{3+}$ | LDB 8 |

*2.2. Detection of security QR codes, bar codes and anti-counterfeiting.*

We have successfully demonstrated the security code detection and anti-counterfeiting of quick response (QR) code and confidential mark on national currency, respectively. In order to investigate the mentioned applications with the use of synthesized LDB4 sample, the grounded phosphor powder was mixed thoroughly with the commercially available polymeric epoxy in the weight ratio of resin:hardener:LDB4-LP phosphor = 2:1:0.5, respectively. This resulted a viscous LP ink used for casting of invisible secret marking in between the pattern of QR code (Meruga et al., 2012). It's to be noted that the QR code used belongs to author's department of the institute and the phosphorescent ink was spread out uniformly using screen printing technique at room temperature (~20°C) and allowed to dry naturally. The developed QR code pattern was excited under white light (D65 lamp) for 15 min and deep blue color luminescence was observed for >4 h when white light source was switched off. Good quality optical images were recorded using blue sensitive digital single-lens reflex Canon camera. Thus, the obtained optical images of secret marking in QR code pattern could be used for auxiliary investigation and detection in dark conditions. Furthermore, LP phosphorescent ink could also be used as an identification tool for the detection of fake currency under ambient light illumination for few seconds and observation in dark box. The invisible secret mark on the currency will be emitting blue color, which is otherwise not possible to observe. This will in turn save the person from the exposure of harmful UV radiations for longer times. Apart from that, synthesized LP phosphor has many more glow-in-dark applications such as rescue sign boards, navigation panel markings in high-rise buildings and highways, textile, luminous paint industries, etc (Kinoshita *et al*., 1999; Terraschke and Wickleder, 2015).

*2.3. Characterization techniques*

The phase analysis of $(Ca_{0.9-x}Si_{0.1})Al_{2-y}O_6:xEu^{2+},yNd^{3+}$ LP phosphors were performed using a X-ray diffraction technique from Rigaku (model: miniflex) having incident radiation characteristics of CuKα (λ = 1.5406 Å) with angle (2θ) ranging $20^o$ to $80^o$. The field-emission scanning electron microscope (model: Zeiss, Supra 40VP) instrument were used for surface morphology observations. The quantitative analysis was done using EDAX (model: Oxford INCA 250) with field emission gun operated at 300 kV. The room-temperature steady-state photoluminescence (PL) studies of LDB-LP phosphor samples were investigated using Horiba Scientific spectrofluorometer (model: Fluorolog-3). The time-resolved PL measurements of the samples were also recorded with the same instrument fitted with microsecond xenon flash lamp as source of excitation and single photon counting system as detector.

**3. Results and discussion**

*3.1. Powder x-ray diffraction (XRD) analysis*

Fig. 1(a) shows the representative optical images of as-prepared LDB4-LP phosphor under white (D65 lamp) and UV (~365 nm) lights, and corresponding deep blue persistent luminescence in dark aroused due to the transitions of $Eu^{2+}$ and $Nd^{3+}$ ions. The as-prepared long persistent $(Ca_{0.9-x}Si_{0.1})Al_{2-y}O_6:xEu^{2+},yNd^{3+}$ (where, x = 0 to 0.05 mol and y = 0 to 0.1 mol) phosphor powders have white body color. The photoluminescence (PL) properties are significantly influenced by the single phase and better crystallinity of LP phosphor. The phase purity of LDB-LP phosphor has been investigated using the X-ray diffraction (XRD) technique. Fig. 1(b) shows XRD pattern of $(Ca_{0.9}Si_{0.1})Al_2O_6$ host lattice with the peaks indexed using PCPDF standard data (#001-0982) of $CaAl_2O_4$ lattice (Ghose, Subrata, Okamura, P. Fujio, Ohashi, 1986; Yang *et al*., 2018). The space group and crystallographic parameters obtained for $(Ca_{0.9}Si_{0.1})Al_2O_6$ phosphor sample was summarized in Table 2. The XRD phase analysis revealed the monoclinic structure having C2/c space group for obtained peaks. It has been noticed that XRD peaks comprises of both $SiO_2$ and $CaAl_2O_4$ phases. However, the phase of $CaAl_2O_6$ LP phosphor was dominant in the structure. Since, the quantity of $SiO_2$ added was nominal (0.1 mol), Si peak was observed in the XRD data with no change in crystalline phase. However, silica content plays an active role as stabilizer for the lattice (Haranath *et al*., 2007; Lee, C. F, Glasser, 1979). This agrees well with the limit of crystallinity of the phosphor powder reports in the monoclinic phase. No diffraction peak associated with impurities was noticed. The lattice parameters were assessed experimentally from observed d-spacing values and (h k l) planes through PC based least square fitting method using "Fullproof software" (Bishnoi *et al*., 2017; Dejene *et al*., 2010; Gupta *et al*., 2015, 2010). It has been observed that the assessed lattice parameters a=9.142Å, b=7.125Å and c=6.568Å are comparable to the standard lattice parameters a=9.609Å, b=8.652Å and c=5.274Å (PCPDF #001-0982). The co-ordination number of $Eu^{2+}/Nd^{3+}$ is found to be seven per unit cell of monoclinic lattice of $(Ca_{0.9-x}Si_{0.1})Al_{2-y}O_6:xEu^{2+},yNd^{3+}$ LP phosphor as per their Wyckoff positions in space group of C2/c. It is observed that the cell parameters and cell volume increases with an increase in the concentration ratio of activator/co-activator ions ($xEu^{2+}/yNd^{3+}$ x=0.01 and y=0.02 mol) and decrease thereafter. Similar features have been observed and deliberated for other rare-earth oxides in our earlier research articles (Haranath *et al*., 2005; Sahai *et al*., 2011; Yerpude *et al*., 2013). The existence of extra quantity of activators/co-activators present in host crystal leads to a situation called lattice contraction effect, which decreases the PL intensities (Gupta *et al.*, 2015; Xu *et al*., 2014). Further details about the lattice contraction effect have been discussed in detail in photoluminescence section. The average particle size of the sample has been estimated using Scherer's formula, by taking different peaks and their full-width at half maximum values were taken into consideration and was found to be ~1 μm. The peaks of monoclinic phase are strongly depends on structural bond length of O-Si, O-Al, Ca-O, Si-Al and Si-Ca to form a compound $(Ca_{0.9}Si_{0.1})Al_2O_6$ host lattice (Jaiswal, Vishnu V. *et al.*, 2020; Kinoshita *et al*., 1999).

*3.2. Surface morphology, structure and compositional analysis*

Scanning electron microscope (SEM) technique was used for determination of surface morphology of the synthesized phosphor. The SEM images of $CaAl_2O_4$ and $(Ca_{0.9}Si_{0.1})Al_2O_6$ LP phosphor are shown at a magnification of 25kX in Figs. 1(c-d), respectively. The observed surface image of $CaAl_2O_4$ was non-uniform, agglomerated irregular shaped structure. However, by the addition of $SiO_2$ to $CaAl_2O_4$ lattice the morphology becomes uniform, textured with significant porosity on the sample surface. It is known that PL intensity will also be affected by different surface morphologies of the phosphor. In order to have enhanced PL intensities along with considerable decay times, phosphor samples with uniform texture and porosities is highly desirable.

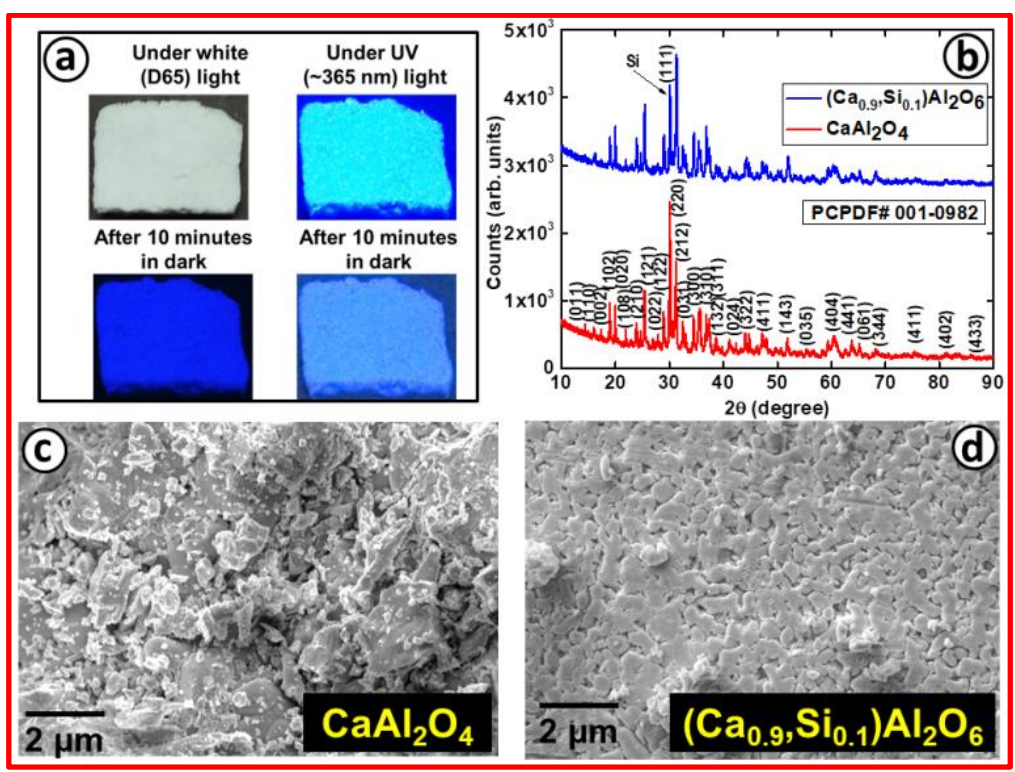

**Fig. 1.** (a) Optical photographs of as-prepared LDB4-LP phosphor kept under white (D65) and under UV (~365 nm) lights; a strong blue persistence emission shown after 10 minutes excitation, respectively (b) XRD patterns (c & d) SEM images of $CaAl_2O_4$ and $(Ca_{0.9}Si_{0.1})Al_2O_6$ phosphors.

**Table 2:** Unit cell and atomic parameters of $(Ca_{0.9-x}Si_{0.1})Al_{2-y}O_6:xEu^{2+},yNd^{3+}$ LP phosphor powder samples have been calculated.

| Unit Cell and Atomic Parameters | |
|---|---|
| Crystal system | Monoclinic |
| Space Group | C2/c |
| V (Å$^3$) | 642.45 |
| $\chi^2$ | 2.15 |
| a | 9.142 Å |
| b | 7.125 Å |
| c | 6.568 Å |
| Alpha (α) | 90° |
| Beta (β) | 90.7° |
| Gamma (γ) | 90° |

To study the microstructure of as-prepared LP phosphor in detail, TEM/HRTEM microscopic analysis has been performed. Figs. 2(a-b) represent the TEM micrographs of LDB4-LP phosphor revealing microstructural information. TEM micrograph of Fig. 2(a) reveals smooth distribution of spherical shaped particles with an average size of ~1 μm. A representative HRTEM image of LDB4-LP phosphor sample is shown in Fig. 2(b). Careful observation of HRTEM image specifies that the sample shows lattice fringes with a interspacing of 0.32 nm, corresponding to the (220) plane of monoclinic phase, which is in agreement with obtained XRD patterns. To probe the elemental composition of the phosphor, EDAX has been performed. The EDAX spectrum shown in Fig. 2(c) confirms the existence of Ca, Si, Al, Eu, Nd, and O in $(Ca_{0.9}Si_{0.1})Al_2O_6:Eu^{2+},Nd^{3+}$ LP phosphor. Since mild reducing atmosphere has been used during preparation, it is understood that Eu and Nd existed in their di and trivalent states. It is apparent from the intensities of the individual elements that the contributions of Nd is significantly more than that of Eu, thereby mitigating the stoichiometry of the as-prepared LDB4-LP phosphor. The histogram shown in Fig. 2(d) represents the presence of each element with its corresponding weight percent.

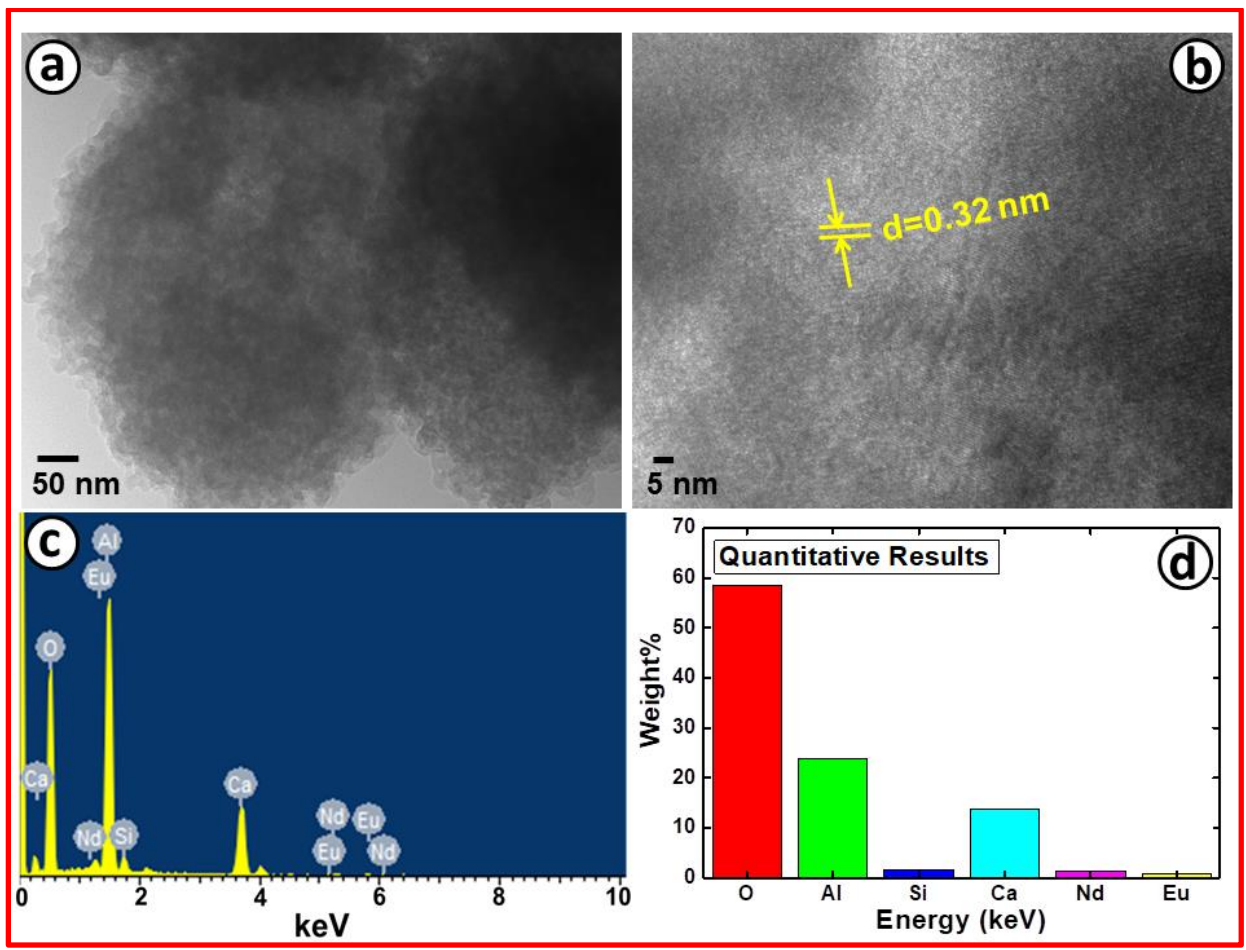

**Fig. 2.** (a) TEM image, (b) HRTEM image showing the lattice fringes, (c) shows the EDAX spectrum and (d) represents the elemental compositions with their corresponding weight percent of LDB4-LP phosphor.

*3.3. Absorption and band gap studies*

The absorption spectra were recorded using UV-Vis-NIR spectrometer for intrinsic $CaAl_2O_4$ and $(Ca_{0.9}Si_{0.1})Al_2O_6$ lattices. Fig. 3(a) shows the absorption spectra of both $CaAl_2O_4$ and $(Ca_{0.9}Si_{0.1})Al_2O_6$ lattices and their energy band gaps were calculated using Tauc relation (Gupta et al., 2015). A drastic shift in the optical band gap values has been observed with a nominal (0.1 mol) addition of $SiO_2$ in $CaAl_2O_4$ compound. A graph has been plotted between $(\alpha h\upsilon)^2$ and $h\upsilon$ (see Fig. 3(b)) and a tangent intercepting the axis containing the photon energy determined the optical band gap. It is evident that addition of $SiO_2$ decreased the optical band gap from 5.64 eV to 4.76 eV. Incorporation of $SiO_2$ in $CaAl_2O_4$ plays significant role in reducing the band gap values that make it more appropriate for increased absorption in the visible region.

*3.4. Photoluminescence excitation (PLE) and emission (PL) characteristics*

It is widely known that the ratio of Eu/Nd affects the distance between Eu-Eu, Eu-Nd, and Nd-Nd ions in the host crystal (LDB samples) (Anesh *et al*., 2014; Swati *et al*., 2015). When the doping concentration of Eu/Nd $(Ca_{0.9-x}Si_{0.1})Al_{2-y}O_6:xEu^{2+},yNd^{3+}$ (where, x = 0 to 0.05 mol and y = 0 to 0.1 mol) and the distance between the Eu-Eu, Eu-Nd, and Nd-Nd ions are large, each ion could be treated as an isolated emission centre and co-centre that emits light independently deprived of any interference. The same has been shown as PL excitation and emission curves of LP phosphor in Fig. 3 (b) and Fig. 3 (c), respectively. Instead, higher doping concentrations of Eu/Nd (beyond x,y=0.01,0.02 mol) could mutually interact with an electric multi-polar process due to the reduced distances between the Eu-Eu, Eu-Nd, and Nd-Nd ions (Gupta *et al*., 2015). In other words, the rate of energy transfer from Eu/Nd ions to the host lattice simply surpasses the radiative rates. Thus, the absorbed photon energy quickly transfers among Eu/Nd ions in the host lattice and decrease the probabilities of the radiative transitions and even decrease the luminescence, if the excited state gets trapped in an energy sink with a high non-radiative deactivation rate constant. This is recognized as a concentration quenching of phosphorescence material (optical photographs depicted in the inset of Fig. 3 (f)) (Bessière *et al*., 2014; Haranath *et al*., 2003; Liu *et al*., 2005; Wang *et al*., 2008). Fig. 3(f) represents the optimization of photoluminescence (PL) intensity due to varying concentration (x,y) of $Eu^{2+}/Nd^{3+}$ from 0 to 0.05 mol and 0 to 0.1 mol in the $(Ca_{0.9-x}Si_{0.1})Al_{2-y}O_6:xEu^{2+},yNd^{3+}$ LP phosphors. Inset of Fig. 3(f) shows the optical images of $(Ca_{0.9-x}Si_{0.1})Al_{2-y}O_6:xEu^{2+},yNd^{3+}$ (where, x = 0 to 0.05 mol and y = 0 to 0.1 mol) LP phosphors excited under ~365 nm. For all the samples, it has been observed that the PL emission peak was centred at 450 nm with a bell-shaped curve. The optimized PL emission spectra show a broad band emission band from 400 to 600 nm with a peak maximum at ~450 nm that could be due to the crystal field splitting of d-orbital associated with $Eu^{2+}$. This in turn indicates that Eu ion is present in the lattice in its divalent state. However, the intensity of PL emission gradually drops down after attaining its maximum intensity at 1 mol% of $Eu^{2+}$ and 2 mol% of $Nd^{3+}$ ion resulting concentration quenching. The maximum PL intensity peak is obtained at 450 nm for LDB4-LP phosphor sample. The PL excitation spectrum of LDB4-LP phosphor was recorded by registering emission at 450 nm, which is shown in Fig. 3(c). This displays a broad band spectrum peaking at 365 nm, corresponding to the crystal field splitting of 5d level in the excited $4f^65d^1$ configuration of $Eu^{2+}$ ions. The observed excitation spectrum of the host lattice in near UV (~365 nm) region endorses the efficient energy transfer from host to $Eu^{2+}$ ions that can gain energy from the host. Moreover, upon visible light excitation, the 4f-5d intra-transitions of $Eu^{2+}$ can happen (Peng *et al.*, 2004). Further, it exemplifies the presence of $Eu^{2+}$ ions and their role as emission centers. This in turn indicates that the as-prepared phosphor can be excited by a variety of light sources, including natural sunlight. Fig. 3(d) shows the PL emission spectrum of LDB4-LP phosphor registered under UV (365 nm) light. This spectrum divulges single symmetric broad-band intense blue emission peaking at 450 nm that could be attributed to the parity-allowed electronic $4f^6 5d^1$ to $4f^7$ transition of $Eu^{2+}$ ions. There is no indication of emission peak corresponding to $Eu^{3+}$ is obtained in the emission spectrum confirming that all $Eu^{3+}$ ions in the sample have been fully reduced to $Eu^{2+}$ ions (Haranath *et al.*, 2007). Here in the current study, $Eu^{2+}$ ions are the luminescent centers and PL arises due to transition from 5d to 4f levels and holes in the traps are accountable for the longer persistence (Clabau *et al.*, 2005). It is well-known that the relative intensities between the broad-band associated with the host (calcium silicoaluminate) and $4f^6 5d^1 \rightarrow 4f^7$ transition of dopant ($Eu^{2+}$ ions) invincibly depends on the activation/deactivation of the host-to-electron-to-trapping centre. Hence, the energy-transfer processes permit the fine-tuning of the chromaticity of the host lattice emission transversely the Commission Internationale d'Eclairage (CIE) diagram. The CIE color coordinates of the samples were calculated from the phosphorescence spectrum using equidistant wavelength method (Jaiswal, Vishnu V. *et al.*, 2020). The CIE color coordinate (x, y) obtained from emission spectra of LDB4-LP phosphor is (0.19, 0.11), is shown in Fig. 3(e). It is located characteristically in the blue region of the chromaticity diagram. Refinement of emission-color across the CIE chromaticity diagram could also be altered by many physico-chemical parameters such as concentrations of $Eu^{2+}$ and/or $Nd^{3+}$; nature of electrons and holes; excitation wavelength, temperature, pH etc. PL kinetic experiment is performed at a fixed excitation wavelength of 365 nm in order to study the PL stability of as-prepared LDB4-LP phosphors (Aitasalo *et al.*, 2006; Bessière *et al.*, 2014; Chang *et al.*, 2006).

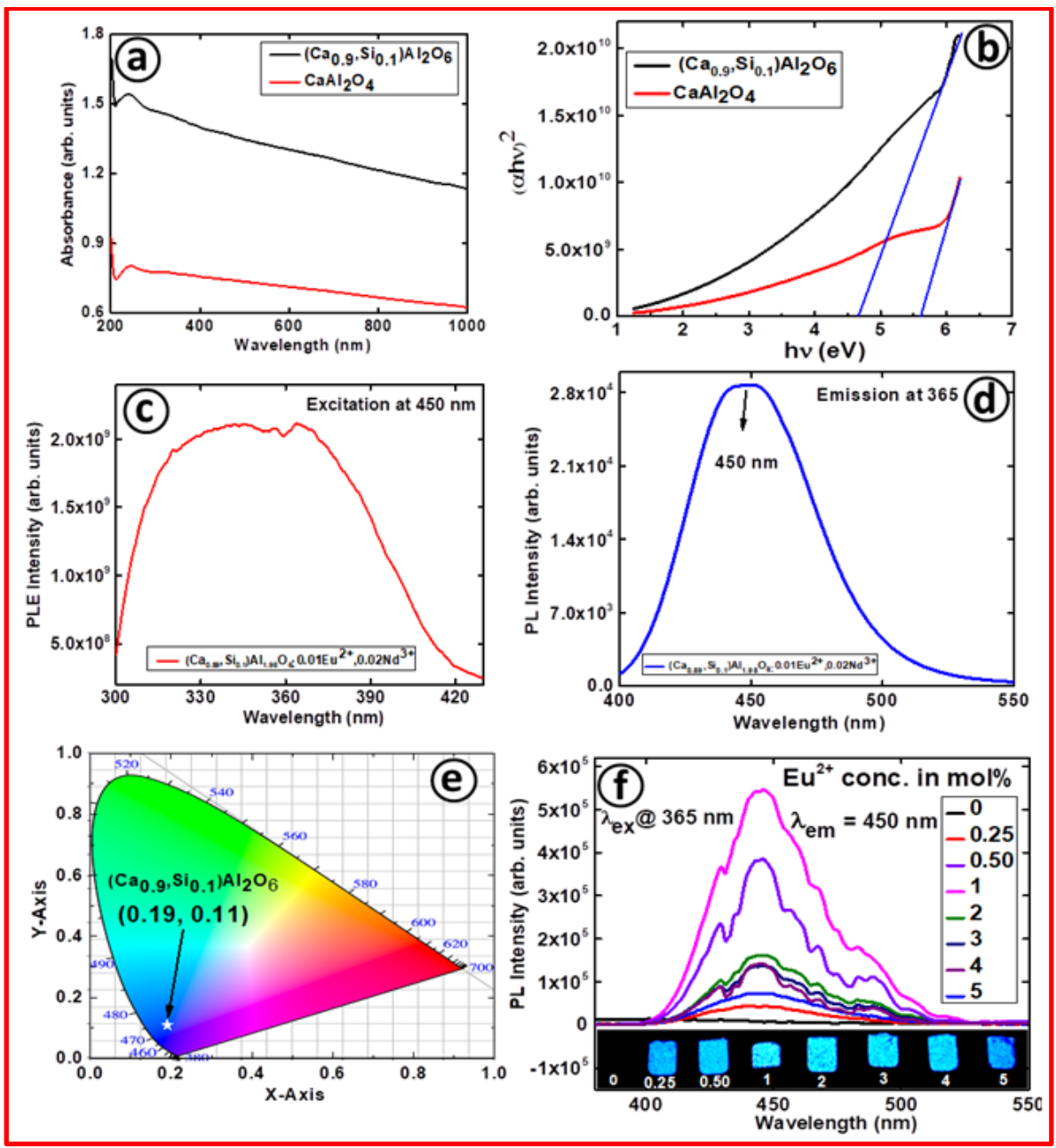

**Fig. 3.** (a) Absorption spectra of $CaAl_2O_4$ and $(Ca_{0.9}Si_{0.1})Al_2O_6$ LP phosphors, (b) plot of $(\alpha h\upsilon)^2$ *vs.* $h\upsilon$ to determine the band gap of intrinsic $CaAl_2O_4$ and $(Ca_{0.9}Si_{0.1})Al_2O_6$ LP phosphors, (c & d) PL excitation and emission spectra of LDB4-LP phosphor recorded at ~450 nm emission and ~365 nm excitation, respectively. (e) chromaticity diagram depicting the color coordinates and (f) PL emission spectra recorded at ~365 nm for optimizing the $Eu^{2+}/Nd^{3+}$ stoichiometric ratio in $(Ca_{0.9-x}Si_{0.1})Al_{2-y}O_6:xEu^{2+},yNd^{3+}$ (where, x = 0 to 0.05 mol and y = 0 to 0.1 mol) LP phosphor. Inset of Fig. 3(f) shows a series of optical images indicating concentration quenching.

*3.5. Time-resolved photoluminescence decay curve*

The time-resolved photoluminescence (TRPL) decay profile of LDB4-LP phosphor was recorded using time correlated single photon counting system and Xenon microsecond flash lamp for 450 nm emission of $Eu^{2+}$ and UV (365 nm) excitation. The TRPL decay curve and subsequent exponential fitting by means of attenuation data parameters of LDB4-LP phosphor are shown in Figs. 4(a-b), respectively. The TRPL decay curve has been plotted on a semi-logarithm paper, in which vertical axis is in log scale. Even after two hours, the kinetic emission intensity of the optimized sample is almost same for the fixed excitation conditions. Tri-exponential decay processes have been observed for LDB4-LP phosphor, in the current study and hence, the TRPL curve has been segmented into three parts. The first part showed that within few seconds of the decay profile, the PL intensity dropped steeply to almost one-tenth of the initial intensity; whereas in the second part, the persistence emission intensity decreased slowly and persisted for tens of minutes and in the third part, there is almost negligible loss of persistence emission intensity that continued for several hours. The TRPL decay profile of LDB4-LP phosphor could be fitted by an empirical equation described here under:

$$I = I_0 + A_1 \exp(-t/\tau_1) + A_2 \exp(-t/\tau_2) + A_3 \exp(-t/\tau_3) \tag{1}$$

where, $I_0$ is the phosphorescence intensity at any time *t* after the excitation source is ceased, $A_1$, $A_2$, and $A_3$ are the corresponding proportionality constants, *t* is the time and $\tau_1$, $\tau_2$, and $\tau_3$ are the decay (attenuation) constants generated from the exponential fit (Gupta *et al*., 2015). The experimentally obtained decay curve shown in Fig. 4(a) is almost satisfying the above equation. Simulating the decay curves and by considering the fitting parameters, the decay constants $\tau_1$, $\tau_2$, and $\tau_3$ have been calculated for the LDB4-LP phosphor. The values thus obtained for τ indicate that there exist three sets of traps leading to three types of decay processes. The value observed for $\tau_3$ is maximum that could be related to deepest trap centres and slowest one in the decay processes. Fig. 4(b) demonstrates the exponential fitting of TRPL decay profile of the LDB4-LP phosphor. The parameters generated from the exponential fitting are $\tau_1$=58.36 min, $\tau_2$=2.67 h and $\tau_3$=4.25 h and are also listed in inset of Fig. 4(b). Considering these values, the average persistent time for LDB4-LP phosphor has been estimated to be more than 4 h. Longer the persistence time, more convenient would be to record, save and protect the information related to security codes and markings. Moreover, one could avoid the unnecessary exposure of harmful UV radiations for the detection of security codes and markings by making use of long persistence phosphorescence emission in a dark box instead of UV box, which is an additional advantage anticipated. Inset of Fig. 4(a) represents the effect of $Eu^{2+}$ concentration on the persistent properties of LDB4-LP phosphors. For all the TRPL curves, the decay time of the persistence emission intensities have been recorded. It has been noticed that the concentration quenching of $Eu^{2+}$ emission begins at x=0.01 mol and the persistence time decreases remarkably beyond this value as shown in inset of Fig. 4(a). The results depicting the effect of $Eu^{2+}$ concentration on the persistence time of LDB4-LP phosphors are in good agreement with the conclusions drawn from Fig. 4(b). The longer persistence time observed is mainly due to the deep trap centers of $Nd^{3+}$ ions that play a substantial role in prolonging the dark persistence for several hours. This signifies the importance of persistence luminescence property of such phosphors for their potential applications in upcoming advanced technologies of our future (Chander and Chawla, 2008).

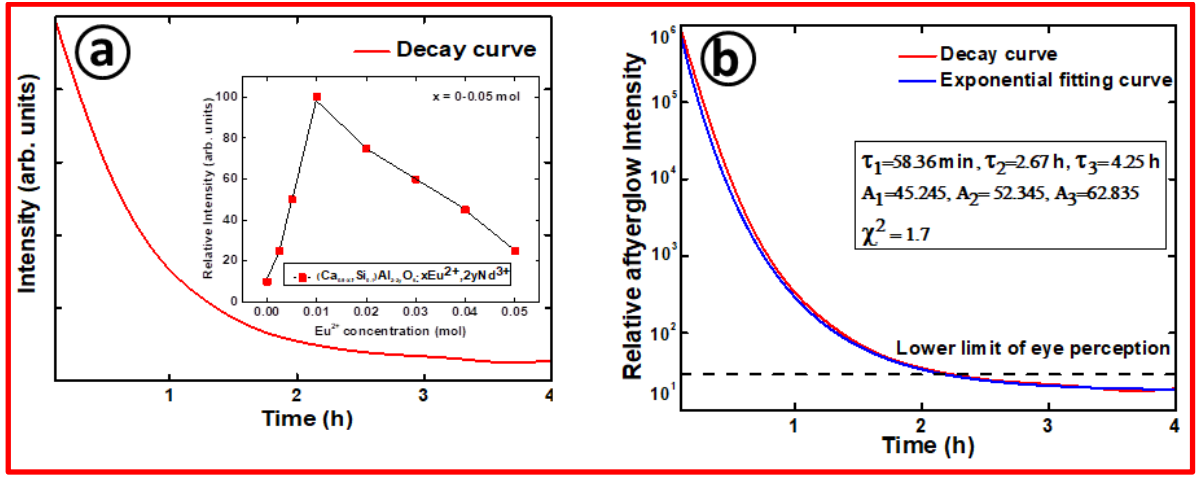

**Fig. 4.** (a) Room temperature TRPL decay curve of LDB4-LP phosphor recorded at ~365 nm and ~450 nm of excitation and emission wavelengths, respectively. Inset of Fig. 4(a) represents the optimal $Eu^{2+}$ activator concentration in $(Ca_{0.9}Si_{0.1})Al_2O_6$ host lattice (b) The exponential fitting of decay curve reveals persistent PL emission that is lasting more than 4 hours and inset shows the attenuation data parameters.

*3.6. Phosphorescent mechanism*

Based on the above discussion, an appropriate mechanism for persistent luminescence and energy transfer model has been projected and depicted in Fig. 5. As mentioned earlier, the mechanism instigating long persistence is due to the hole trapping–transporting–detrapping phenomena (Yamamoto and Matsuzawa, 1997). When LDB4-LP phosphor is excited by UV (~365 nm) light, a more number of charge carriers are produced due to the presence of Si in the LDB4-LP phosphor host lattice, as a consequence reduction in band gap and direct excitation of $Eu^{2+}$ ions arises. A sufficient number of positive holes are created in the valance band and certain free holes are captured by $Nd^{3+}$ ions that act as hole traps (Dorenbos, 2005; Haranath *et al.*, 2007). Once the source of excitation is removed, the holes present at defect sites of $Nd^{3+}$ absorb the pervading thermal energy and get de-trapped to the ground state of $Eu^{2+}$ with strong blue (450 nm) emission. It is evident that the depth and the density of $Nd^{3+}$ hole traps increase the persistence time period significantly. Thus, it could be understood that the trapping and de-trapping of holes by the trap levels play a vital part in deciding the long persistence behaviour in LDB4-LP phosphors.

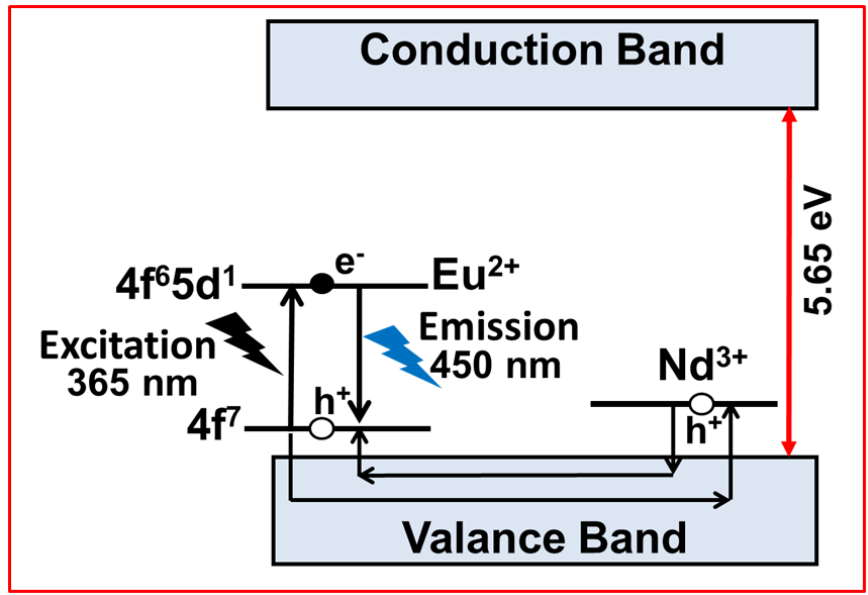

**Fig. 5.** Proposed phosphorescent mechanism of LDB4-LP phosphor, $e^-$ and $h^+$ denotes electron and hole, respectively.

## 4. Strategic applications of LDB4-LP phosphor sample

*4.1. Knocking out the fake security codes and combat counterfeiting*

The potential applications of LDB4-LP phosphor was validated for security code applications involving quick response (QR) code, bar code and anti-counterfeiting of currency notes, which is not published previously for this phosphor. We have formulated a novel LDB4-LP phosphor ink by dispersing the finely grinded phosphor powder sample in epoxy resin in ratio of 0.5:2 by weight, respectively. The stability of the phosphorescent ink was found to be strongly dependent on the process of mixing, temperature, relative humidity, pH and concentration of the powder and epoxy. Hence, the process of mixing must be carried out at room temperature ($20^oC$) for not less than 3 h duration in a dust-free and dry environment to produce a stable colloidal suspension. Fig. 6 illustrates the making of LDB4-LP phosphor ink and its visualization under white light, UV (365 nm) and dark conditions for various strategic applications.

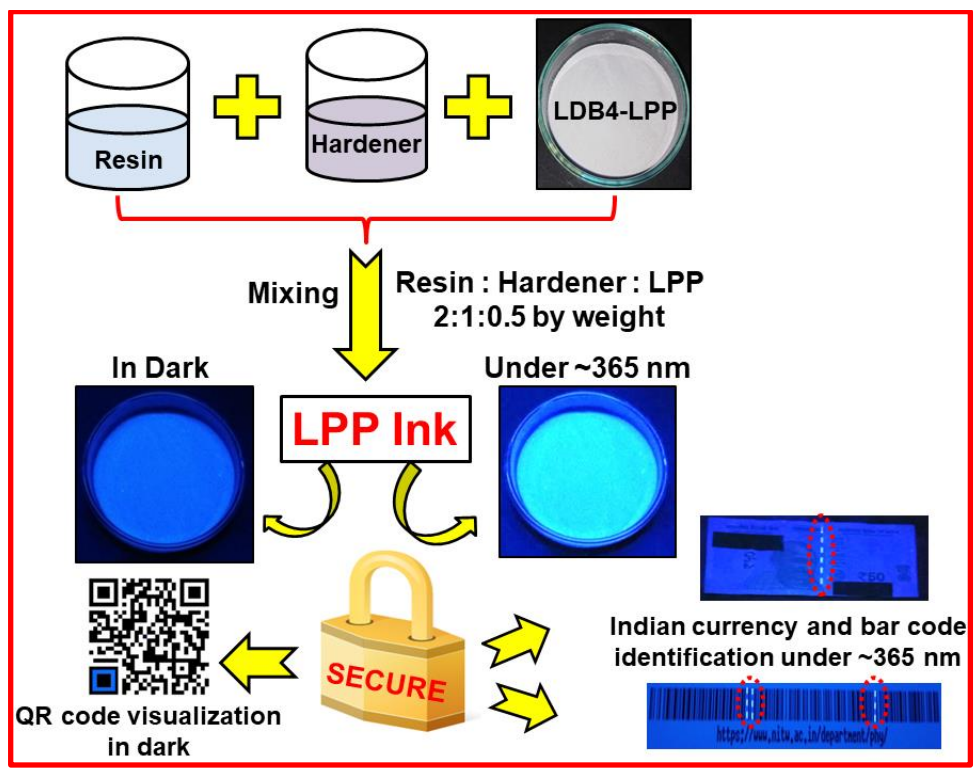

**Fig. 6.** Formulation of LDB4-LP phosphorescent ink and its potential applications.

*4.2 QR and Barcode detection*

Fig.7. shows a representative quick response (QR) code having unique black and white pattern. In every code, there is some hidden information about a specific website and status of error correction level (Meruga *et al*., 2012). One more unique feature of the QR code is the use of an encoding disguise over the content within. The active area of each QR code depends on the variety of mask used and the information to be stored (Meruga *et al*., 2015, 2012). Elimination of huge portions of darkened modules or extremely repeating white modules gives the guarantee of readability of the text. There will be three square boxes located at the bottom and top left side, and top right side corners that serve as anchors and safeguard configuration for fast readability. Depending upon the version used for designing QR code, the smaller dark blocks are distributed throughout the code beginning from the bottom right side corner (Krombholz *et al*., 2014). Interestingly, a row of dark modules between the top left to the top right anchors, as well as a column of alternating black and white modules between the top left and the bottom left anchors indicate the timing pattern. Information about the version used will be located above the bottom left and to the left of the top right anchors. Information related to formatting will be located around the top left to the right of the bottom left anchors, and below the top right anchor. These are the elementary properties that assist us to identify, classify, and standardize QR codes (Krombholz *et al*., 2014; Law and So, 2010). Although QR codes have significantly more storage capacity than their 1-D counterparts, the storage capacity of traditional mainstream QR codes is limited to a few thousand alphanumeric characters. The general use of QR codes now-a-days is for advertising purposes (generally URL listings for institute/company websites) to provide additional information to consumers about a given product or service (Sourceforge, 2020). In the current paper we present techniques to identify and secure the multi-layered QR codes. Secret codes are prepared such that they are invisible under room light when viewed by naked eye. In the case of QR codes, the printed lines and blocks of the QR code are invisible under room light. Fig.7 represents the optical images of representative QR codes that are excited under white (D65) light and viewed under extreme dark condition for the detection of confidential marked codes using LDB4-LP phosphor. Hence, long persistent blue (~450 nm) PL emitted from the coding channels of QR code secure the data, information and storage capacities.

Typical bar code used in labeling and pricing a specific product in the market provide a simple and most convenient way of product identification in the stock. Barcodes are usually classified as uni-dimensional (1-D) and bi-dimensional (2-D). The 1-D barcodes use lines and spaces with different widths that signify 'product identification', whereas, 2-D barcodes use symbols and accentuate 'product descriptions'. 2-D barcodes are superior to 1-D barcodes as they can accommodate supplementary storage capacity, security, and readability features (Hanna and Pantanowitz, 2016; Liu *et al*., 2020). However, additional facet of security could be introduced for 2-D barcodes using LDB4-LP phosphor under the excitation of UV (365 nm) light.

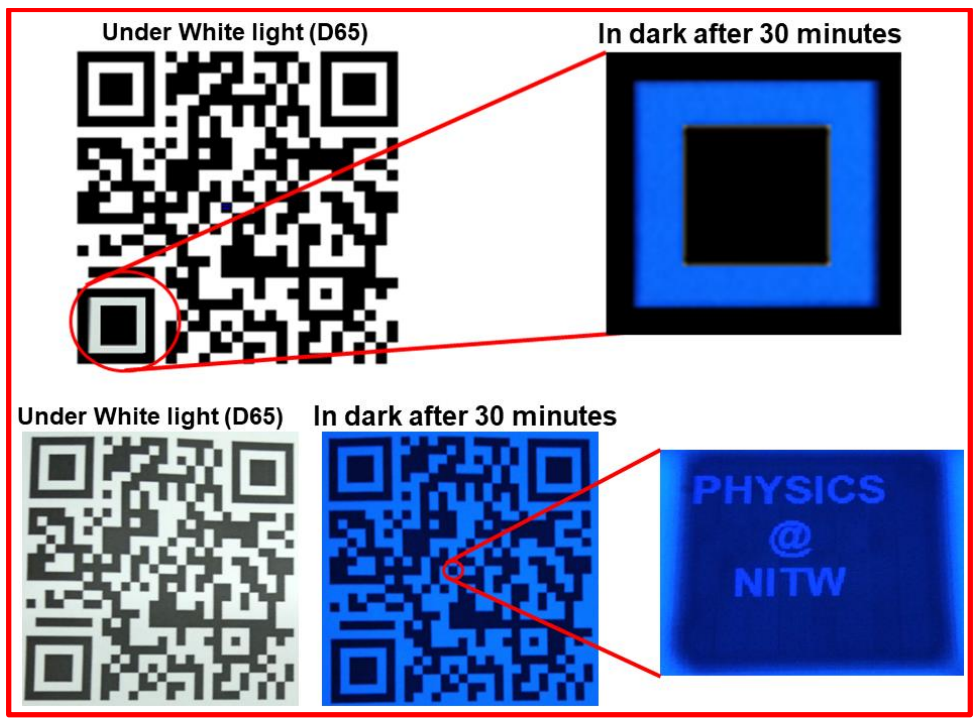

**Fig. 7.** Optical images of QR codes that are excited under white (D65) light and in extreme dark condition for the visualization of lower left anchor as well as hidden text with blue (450 nm) phosphorescence from LDB4-LP phosphor ink.

*4.3 Anti-counterfeiting*

Counterfeiting has been considered as one of the major risk for global economy and already it has crossed the alarming level in G20 countries and affected about 10% of world trade (Kanika *et al*., 2017). According to the 2019-report of the Organization for Economic Co-operation and Development (OECD) around $650,000 million global economy was affected by counterfeiting (OECD/EUIPO, 2019; Organisation for Economic Co-operation and Development & European Union Intellectual Property Office, 2017). In recent times, various organic and inorganic based luminescent materials, fluorescent nanomaterials, semiconductor quantum dots (QDs), carbon QDs, plasmonic nanomaterials, etc. were used to address this issue. However, significant improvement has not been achieved in eradicating this problem completely (Li, *et al.,* 2013). In the current paper, we have explored a blue (~450 nm) emitting LP phosphor ink to serve the purpose of anti-counterfeiting applications on various templates as shown in Fig. 8(a-b). To show the practical utilization of this security feature, a representative student identity card of the host institute (National Institute of Technology Warangal) and Indian currency have been inscribed with an invisible secret barcode strip, which could be seen only under UV (365 nm)

light as shown in Fig. 8(a). The hidden blue (~450 nm) emitting broken line marked by the phosphorescent ink is visible with better contrast when placed under a dark box. This novel way of visualizing invisible mark/strip will certainly combat counterfeiting and address the issue of fake barcodes on any consumer product or machinery. Moreover, this technique does not require any sophisticated instrumentation to eliminate the problem of counterfeiting globally.

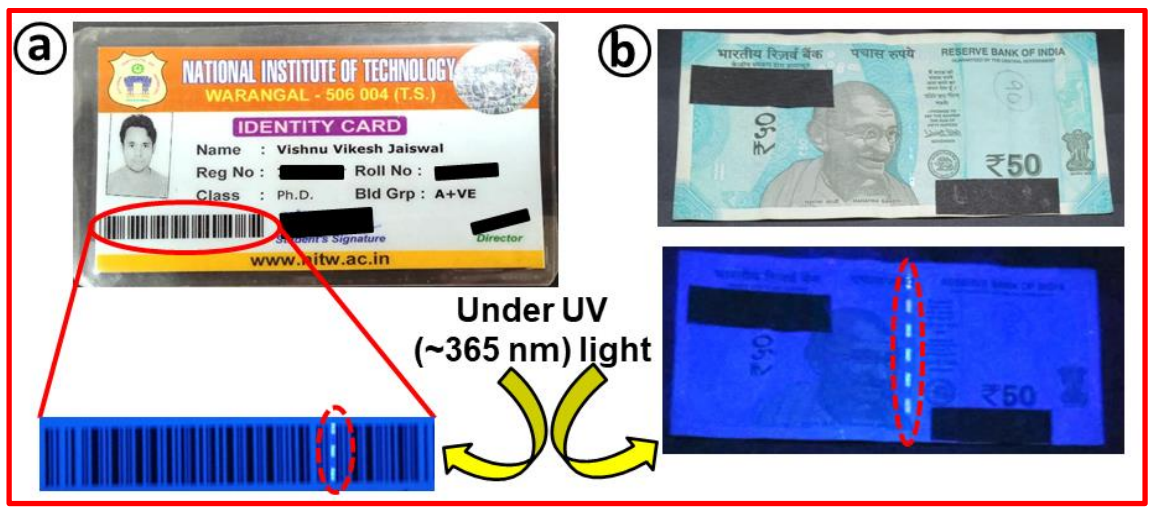

**Fig. 8.** (a) The optical images of student ID card and (b) Indian currency have been protected by identifying the secret codes under the exposure of UV light of ~365 nm using LDB4-LP phosphor.

## 5. Conclusions

In conclusion, a novel class of $(Ca_{0.9}Si_{0.1})Al_2O_6:Eu^{2+},Nd^{3+}$ long persistent phosphor was synthesized and formulated into a phosphorescent ink to potentially use for knocking out fake security codes and combat counterfeiting of currency notes. The phosphorescent ink is of a special kind that absorbs and stores the energy from ultra violet light, white light and natural sunlight, and emits efficient blue (~450 nm) photoluminescence noticeable for many hours under dark conditions. However, the optimum PL emission centred at ~450 nm has been registered for UV (365 nm) excitation, which is due to the transitions from $4f^6 5d^1$ to $4f^7$ energy levels of $Eu^{2+}$ ions. The fundamental mechanism for long persistence produced was due to trapping and de-trapping of holes at $Nd^{3+}$ sites situated adjacent to the valance band. For the optimized stoichiometric composition of LDB4-LP phosphor, the persistence lasted for >4 hours for dark adapted human eye. Additionally, the phenomenon of long persistence empowers to make unique hidden markings on consumer products and certificates using quick response (QR) code patterns. Methodical studies have been performed on various templates kept under UV (~365 nm) excitation to identify fake currencies, barcodes and combat counterfeiting.

## Acknowledgements

The authors wish to thank the Director, NIT Warangal for extending the necessary laboratory facilities to carry out the research work. One of the authors (VVJ) is grateful to the Council of Scientific & Industrial Research (CSIR), Government of India for providing the financial support under CSIR-Senior Research Fellowship [No. #09/922(0007)/2018-EMR-1].